%% file: paper.tex
\documentclass[10pt,preprint]{sigplanconf}
\usepackage{times}

\usepackage{datetime}
\usepackage{url}
\usepackage{hyperref}
\usepackage{graphicx}
\usepackage{listings}

\setcitestyle{numbers,open={[},close={]}}

\newcommand{\SysName}[0]{{Quest}}
\newcommand{\SysNameS}[0]{{quest}}
\newcommand{\VirtName}[0]{{Quest-V}}

\date{}

\authorinfo{
\normalsize Ying Ye, Zhuoqun Cheng, Soham Sinha, Richard West
  \\ Computer Science Department \\ Boston University ~\\
  Email: \{yingy,czq,soham1,richwest\}@cs.bu.edu
}

\begin{document}

\title{vLibOS: Babysitting OS Evolution with a Virtualized Library OS} 
\maketitle

\begin{abstract} 
Many applications have service requirements that are not easily met by
existing operating systems. Real-time and security-critical tasks, for
example, often require custom OSes to meet their needs. However, development
of special purpose OSes is a time-consuming and difficult exercise.  Drivers,
libraries and applications have to be written from scratch or ported from
existing sources. Many researchers have tackled this problem by developing
ways to extend existing systems with application-specific services. However,
it is often difficult to ensure an adequate degree of separation between
legacy and new services, especially when security and timing requirements are
at stake.  Virtualization, for example, supports logical isolation of separate
guest services, but suffers from inadequate temporal isolation of
time-critical code required for real-time systems. This paper presents vLibOS,
a master-slave paradigm for new systems, whose services are built on legacy
code that is temporally and spatially isolated in separate VM
domains. Existing OSes are treated as sandboxed libraries, providing legacy
services that are requested by inter-VM calls, which execute with the time
budget of the caller. We evaluate a real-time implementation of
vLibOS. Empirical results show that vLibOS achieves as much as a 50\%
reduction in performance slowdown for real-time threads, when competing for a
shared memory bus with a Linux VM.
\end{abstract}

\section{Introduction}
\input{introduction}

\section{Motivation}
\label{sec:motiv}
\input{motiv}

\section{vLibOS Design}
\label{sec:design}
\input{libos}

\section{Implementation: A Multicore Real-Time System}
\label{sec:impl}
\input{case}

\section{Evaluation}
\label{sec:eval}
\input{eval}

\section{Related Work}
\label{sec:related}
\input{related}

\section{Conclusions and Future Work}
\label{sec:con}

This paper presents vLibOS, a master-slave paradigm that integrates services
from multiple OSes into a single, custom system. The approach allows
pre-existing OSes to provide legacy services to new systems with specialized
QoS requirements. The new system features are implemented in a master OS that
calls upon legacy system software in different virtual machines. We argue that
the master OS should manage shared hardware resources on the platform for
meeting its targeted QoS. As it is able to coordinate and schedule the
execution of services in other VMs, inter-VM performance isolation is greatly
improved. This is critical to systems that require temporal predictability
(e.g., real-time guarantees).

In our prototype system, \VirtName{}, we developed a partitioning hypervisor
with vLibOS API support. The proposed unified scheduling mechanism enables
legacy services to be an extension of a client thread in the master OS, thus
granting the control of a Linux VM to the master. The master, \SysName{}, then
uses a latency-based memory throttling technique to regulate the shared memory
bus. This avoids excessive concurrent memory accesses from separate cores that
would otherwise lead to unpredictable execution times of real-time
services. Our experiments show the benefits of the vLibOS system design.

Future work will investigate the benefits and tradeoffs of the vLibOS approach
for a diverse range of applications with timing, safety and security
requirements (e.g., smart Internet-of-Things, cloud systems). We will also work
on ways to improve the utilization of cores in a vLibOS system.


\bibliography{reference}

\end{document}

%% file: introduction.tex
Computer hardware is evolving rapidly today, with computing devices now
supporting multiple cores, virtualization, and general purpose graphics
processing units (GPGPUs). Modern hardware capabilities have led to the
emergence of new classes of applications requiring safety, security and timing
predictability. For example, driverless cars, unmanned aerial vehicles, smart
manufacturing devices and other Internet-of-Things (IoT) applications require
data to be exchanged and processed both securely and predictably, and
decisions to be made in real-time. However, existing general-purpose OSes
(GPOS) are ill-equipped to address the needs of these emerging applications;
many existing systems are not easily extensible due to their monolithic
design that does not securely or safely isolate new features. Similarly,
general-purpose systems focus on fair resource management for multiple users,
rather than timing guarantees required of certain mission-critical
applications.  This means developers are either required to apply ad hoc
patches to existing systems, or they must write operating systems from scratch
for their specific application needs. Patching existing systems does not
easily achieve the desired end goals because of the fundamental mismatch
between a GPOS' design goals and custom applications'
requirements. Alternatively, writing a new OS is a time-consuming and
difficult exercise. Device drivers, libraries and application-programming
interfaces must all be written for a new OS to support any kind of non-trivial
application.

Traditional wisdom~\cite{Ford:97} argues that a modular design makes kernel
writing easier. Common features are made available for reuse as kernel modules,
as long as module interfaces are well-defined. Legacy device drivers can be
ported to a new kernel by encapsulating them within adaptation layers.
However, strict modularization tends to involve performance tradeoffs and
creating adaptation layers is often a difficult task.

\begin{figure}
\centering
\includegraphics[scale=0.48]{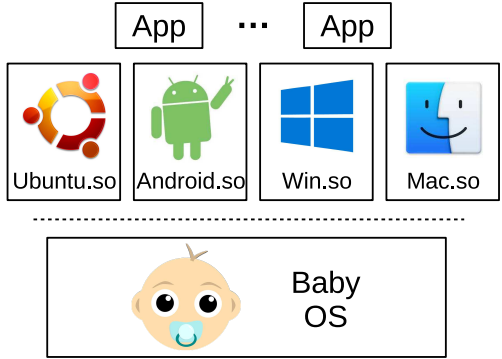}
\caption{vLibOS Concept}
\label{fig:overview}
\vspace{-0.15in}
\end{figure}

Using virtualization extensions found on many of today's multicore
processors, it is possible to treat existing systems as guest services in the
development of new OS functionality. A new OS implemented in one guest domain
leverages legacy OS features in another guest domain, without requiring a
clean-slate implementation of everything. A virtual machine (VM) provides a
sandbox for OS developers to implement innovative new mechanisms, while
delegating certain services to legacy OSes~\cite{Ammons:07}. However, VMs
running on the same processor compete for shared hardware resources such as the
last-level cache (LLC) and the memory bus~\cite{Ye:14, Ye:16}. Without careful
management of VM execution, system performance is arbitrarily impacted.

We attribute the interference issue to the lack of a global resource manager.
In a virtualized environment, the hypervisor assumes the role of a shared
resource manager. Nevertheless, we have discovered several issues with this
approach, which will be discussed in Section~\ref{sec:motiv}. In this paper,
we argue the case for a newly-developed {\em baby OS} to act as a global
resource manager for OS services spanning multiple VMs. This way, the baby OS
in one VM is able to control the resource usage of legacy OSes in other VMs,
ensuring the desired behavior of the overall system. For example, by
regulating DRAM traffic from each core and reducing bus contention, the
Quality-of-Service (QoS) for timing-sensitive applications is greatly
improved~\cite{Yun:13}. In general, Denial-of-Service (DoS) attacks on shared
resources~\cite{Moscibroda:07} from a compromised legacy OS VM are
preventable.

In this paper, we describe a novel system architecture
(Figure~\ref{fig:overview}) for building QoS-aware OSes. The master-slave
paradigm is adopted, in which the baby OS is the master and legacy OSes act as
slaves. With the assistance of hardware virtualization, slaves are controlled
by the master, to provide legacy services, including libraries, device
drivers, and system calls. We call each slave a virtualized library OS, or
{\em vLib OS}. A vLib OS is not a traditional library OS. Traditional library
OS models focus on re-writing OS services as application components, often
requiring significant engineering effort. On the contrary, our vLibOS model
helps the construction and adoption of new OSes by relying on legacy OSes to
provide a feature-rich environment with minimum effort. Moreover, unlike
traditional library OS designs that treat an OS as a set of application
libraries, we view an entire legacy OS as a single library to be added to the
master. Thus, a vLib OS acts as a standard library whose execution is under
the control of the master. With this extended control, the master is able to
manage global hardware resources, providing enhanced QoS.

The contributions of this work are threefold:

\begin{list}
{$\bullet$}
{
\setlength{\topsep}{\parskip}
\setlength{\parsep}{0in}
\setlength{\itemsep}{0in}
\setlength{\parskip}{0in}
}

\item We propose the vLibOS model for building new OSes, with legacy support
fulfilled by existing OSes. We argue that shared hardware resources should be
managed by the baby OS instead of the hypervisor.

\item We introduce an implementation of vLibOS to ease the development of a
feature-rich real-time system.

\item We also discuss the application of the vLibOS model and its alternative
implementations in different settings.

\end{list}

The rest of this paper is organized as follows: The next section describes
motivation and challenges for the use of virtualization in OS evolution.
Section~\ref{sec:design} discusses the vLibOS model and its general design
principles. A real-time system implementation of vLibOS is then shown in
Section~\ref{sec:impl}, along with a discussion of the pros and cons.
Section~\ref{sec:eval} describes an evaluation of our real-time system and how
it meets predictable execution requirements. This is followed by a description
of related work in Section~\ref{sec:related}. Finally, conclusions and future
work are outlined in Section~\ref{sec:con}.

%% file: motiv.tex
Operating systems have been evolving for decades, in response to hardware
advances and new applications. The addition of new mechanisms and policies
oftentimes demands a substantial restructuring to the system. For example,
replacing the O(1) scheduler in Linux with the Completely Fair Scheduler (CFS)
required an entire overhaul of the scheduling infrastructure~\cite{linux_cfs}.
In other cases, it is almost impossible to extend the functionality of an
existing system without completely rewriting fundamental components. For
example, adding real-time guarantees to a previously non-real-time OS,
improving scalability beyond a previous system limit, or heightening system
security. Under these situations, a new kernel~\cite{Liedtke:95, Baumann:09,
Engler:95} needs to be developed.

One of the biggest stumbling blocks in the construction and adoption of a new
OS is the support for legacy applications, libraries and device drivers.
Millions of applications are immediately available on existing OSes while
thousands more are being developed every day. Similarly, third parties
regularly contribute libraries and device drivers to pre-existing systems such
as Windows and Linux. In contrast, a new OS might only support a few features,
and the effort required to port the rest is potentially many man-years.

Virtualization provides an opportunity to combine legacy system features with
new OS abstractions, greatly saving engineering cost. It is possible to
encapsulate full-blown OSes in separate virtual machines, and have their
services made accessible to another OS using an appropriately implemented
Remote Procedure Call (RPC) mechanism. 

Consider, for example, the development of a mixed-criticality
system~\cite{Vestal07}, combining components with different timing, safety and
security criticality levels~\cite{integrity178b}. Here, criticality is defined
in terms of the consequences of a system or component failure. Such a system
might be applicable in the automotive domain, where the services for chassis
(e.g., braking and stability control), body (e.g., lighting and climate
control), and information (e.g., infotainment and navigation) are consolidated
onto a single multicore platform. Such a system might employ a legacy Android
system to support low-criticality infotainment and navigation services, while
implementing high-criticality services like managing vehicle stability and
braking as part of a new AUTOSAR-compliant~\cite{autosar} OS. The
high-criticality services require temporal and spatial isolation from legacy
code to ensure their correct timing behavior. Moreover, the failure of a
low-criticality service should not compromise the behavior of one with higher
criticality.

While virtualization provides a way to integrate legacy functionality into a
new OS, it does not solve the problem of performance isolation. Virtualization
has traditionally only provided a logical separation between guest virtual
machines. However, when multiple VMs execute on separate cores, they compete
for shared last-level caches, memory buses and DRAM~\cite{Ye:16, Yun:14}.
Uncontrolled access to shared physical resources leads to detrimental
performance interference. Newly-developed timing-sensitive services are in
jeopardy of unbounded timing delays from the execution of other VM-based
services.

It is possible to modify a hypervisor in existing virtualization systems such
as Xen or VMware ESXi to ensure performance isolation. This requires the
implementation of timing-aware resource management policies. However, there
are several issues with this approach. First, since every guest OS already has
its own mechanisms and policies for managing shared resources, adding
OS-specific policies to a hypervisor not only creates unwanted
inter-dependencies but also duplicates functionality. Modifications to a guest
OS may require alterations to the underlying virtualization infrastructure as
well.  Second, adding resource management policies to the hypervisor increases
the size of the Trusted Computing Base (TCB), potentially reducing the
security and reliability of the entire system~\cite{Szefer:11}. Third,
hypervisors perform resource accounting at the granularity of virtual CPUs,
which adds overhead to the accounting mechanisms already in the guest OS. For
instance, suppose a QoS-aware guest OS tries to improve worst case DRAM access
latency, by tracking every thread’s DRAM access rate and performing traffic
regulation when the memory bus is heavily contended~\cite{Yun:13}; to avoid
inter-VM interference and meet the required QoS, the hypervisor also needs to
manage the DRAM access of every VM, introducing an extra layer of resource
accounting on the bus. Fourth, applications that request services from other
guest OSes, henceforth referred to as {\em dual-mode} applications, require
resource accounting (e.g., CPU budget, cache occupancy, memory bandwidth) to
span multiple VMs.  Current hypervisors only account for resource usage by
individual VMs. Without accurately accounting resource usage for dual-mode
(inter-VM) applications, resource management cannot be carried out
effectively.

With the above considerations in mind, we believe it is still possible to use
virtualization for running legacy services in conjunction with newly-defined OS
functionality. However, for virtualization to be effectively used in the
construction of evolvable and QoS-aware systems, it is important to provide
performance isolation and proper resource accounting across multiple VMs. This
requires new hypervisor technologies that will be described in the next
section.

%% file: libos.tex
\begin{figure}
\centering
\includegraphics[scale=0.43]{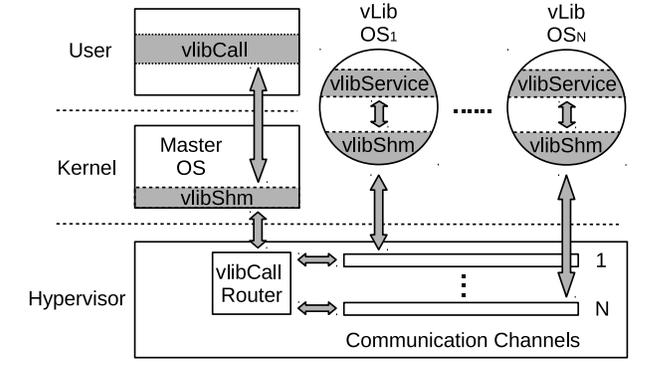}
\caption{vLibOS Architecture}
\label{fig:arch}
\end{figure}

The architecture of vLibOS is shown in Figure~\ref{fig:arch}. It consists of a
set of user APIs, a master OS, a set of vLib OSes, and a hypervisor. The master
OS implements new features while leveraging pre-existing services in one or
more vLib OSes. Each vLib OS runs under the control of the master.

\subsection{User APIs}
\label{sec:api}

Each vLib OS runs a server program to provide services to the master. A call to
\begin{center} {\em channelAddr* vLib\_listen(port)} \end{center}
from the {\em vlibService} library blocks the entire vLib OS until it is
requested to execute on behalf of a service caller. This causes all virtual
CPUs to be suspended inside the hypervisor, waiting for vLib calls from client
threads inside the master OS. A port number is used to uniquely identify vLib
OSes. Once this function is unblocked, it returns a virtual address to the
communication channel established between the client and itself, which contains
all the data needed for the service. The data includes the function requested,
the name of the library that contains the function, and the input data.
Service completion causes the server to write the output back to the channel,
followed by calling {\em vLib\_listen} again. This leads to a completion signal
being sent to the client, and the vLib OS waits for another request. An example
vLib server is listed below (Listing~\ref{code:server}).

\footnotesize
\begin{lstlisting}[caption={Example vLib Server}, label=code:server]
while ( channel = vLib_listen(port) ) {
    /* locate service */
    /* unmarshal data */
    /* perform service */
    /* write result back to channel */
}
\end{lstlisting}
\normalsize

A client application uses the {\em vlibCall} library to make vLib calls into
services from one of the vLib OSes. The following APIs are provided:

\begin{list}
{$\bullet$}
{
\setlength{\topsep}{\parskip}
\setlength{\parsep}{0in}
\setlength{\itemsep}{0in}
\setlength{\parskip}{0in}
}
\item {\em errCode vLib\_init(port, channel\_size, **channel\_addr);} 
\item {\em errCode vLib\_call(port, timeout);}
\item {\em errCode vLib\_async\_call(port, callback, timeout);}
\item {\em errCode vLib\_channel\_destroy(port);}
\end{list}

{\em vLib\_init} establishes a communication channel to the vLib OS listening
on {\em port}. After data is copied into the channel, a subsequent {\em
vLib\_call} requests a service with an optional {\em timeout} for the
server-side processing. Multiple service requests to a single vLib OS are
serialized in FIFO order. The timeout is used to terminate the service wait,
which guarantees a bounded delay for the call and avoids liability
inversion~\cite{Hand:05}. A {\em vLib\_call} is a blocking (i.e., synchronous)
request, while a {\em vLib\_async\_call} provides a non-blocking asynchronous
interface. An existing channel is closed and its resources are reclaimed
through {\em vLib\_channel\_destroy}.

\subsection{Master OS}

A vLibOS system includes a single master OS, which acts as a centralized
manager of all hardware resources with the help of a hypervisor. Both native
applications and {\em dual-mode} applications (spanning the master and one or
more vLib OSes) are supported. Dual-mode applications start in the master, for
proper cross-VM resource accounting. A {\em vlibShm} kernel module maps
communication channels into applications' address spaces.

\subsection{vLib OS}

A vLib OS is any existing OS, such as a UNIX-based system with process address
spaces, or a traditional library OS having a single address
space~\cite{Madhavapeddy:13}. Each vLib OS provides functionality for use by
the master OS. The hardware Performance Monitoring Unit (PMU) is virtualized to
a vLib OS. Core-local performance counters will not be exposed if being used by
the hypervisor/master. Also, global performance events are made inaccessible.
This hardens security isolation between the master and vLib OSes. For example,
information leakage from side channels based on PMU data is
avoided~\cite{Kocher:99}. The separation of a master OS and one or more vLib
OSes provides the basis for a mixed-criticality system. Each vLib OS
establishes a sandbox domain for services of different timing, safety and
security criticalities. As with the master OS, each vLib OS uses a {\em
vlibShm} module to map communication channels into the server's address space.

\subsection{Hypervisor}

The hypervisor in vLibOS is responsible for booting OSes and delegating
resources (CPUs, memory and devices) to them. It provides an interface to VMs
to support the user APIs from Section~\ref{sec:api}. For vLib calls, it
allocates communication channels between client applications and servers upon
request, and routes calls to the right destinations (using the vLibCall Router
in Figure~\ref{fig:arch}). To avoid time-related issues inside a vLib OS due to
blocking, guest time is virtualized. More importantly, the hypervisor empowers
a master OS with the capability to block and wake up other VMs. This means the
execution of a vLib OS is integrated into the scheduling framework of the
master, for effective hardware resource management.

\begin{figure}
\centering
\includegraphics[scale=0.28]{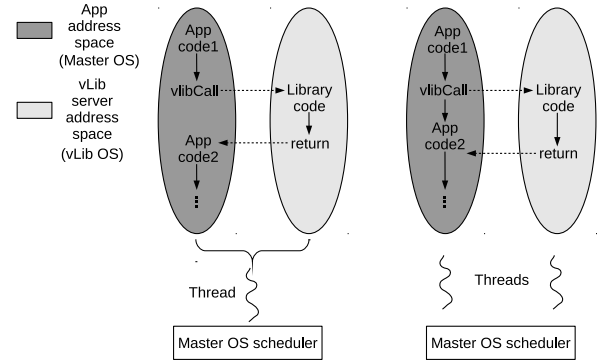}
\caption{vLibOS Unified Scheduling (left is sync call, right is async call)}
\label{fig:sched}
\end{figure}

Figure~\ref{fig:sched} illustrates the unified scheduling mechanism in vLibOS.
A vLib call resembles an RPC, with the caller and callee occupying separate
address spaces. However, for synchronous vLib calls, the callee shares the
same resource accounting entity (thread from the master OS) with the
caller. The callee in a vLib OS executes with the CPU budget of the calling
thread from the master. Once the budget is depleted, or preemption occurs in
the master, the callee is descheduled. For asynchronous vLib calls, the master
OS manages the callee as a second thread. This mechanism extends the
capability of a resource-aware scheduler in the master OS, for managing
resource contention across the entire platform.

\subsection{Applications}

Applications can utilize vLib OS services in several ways. In the first case,
dual-mode applications start in the master and send each request through a vLib
call. As a vLib call incurs higher overhead than a library call, this approach
should be avoided on performance critical paths. In the second case, a
dual-mode application first makes an async vLib call (with no input) to pass
its CPU budget to a vLib OS. Then it writes a series of service requests to the
communication channel. The server on the other side, upon returning from {\em
vLib\_listen()}, starts polling on the communication channel to get the actual
requests. When an ending signal is received, the server jumps back to {\em
vLib\_listen()}, indicating the end of this async vLib call session. This
approach, while greatly reducing system overhead for sending each request,
locks a vLib OS for a longer period as well, thus blocking other applications
from requesting services. In the third case, applications run entirely inside a
vLib OS, while a dummy thread is created in the master. The dummy thread makes
a single sync vLib call with no input. After receiving the call, the server
terminates without signaling service completion. Consequently, all other
applications on the vLib OS inherit the dummy thread's CPU budget and continue
execution.

%% file: case.tex
We implemented our vLibOS architecture by extending an existing virtualization
system, referred to as \VirtName{}~\cite{Li:14}. \VirtName{} is targeted at secure and
predictable embedded systems, where virtualization is used to isolate legacy
functionality from timing, safety and security-critical custom OS
features. The system currently runs on the x86 (IA32) architecture with VT-x
virtualization extensions. It provides separate VM domains for a master OS and
a Linux vLib OS. Our master OS, \SysName{}~\cite{Ye:16}, is a
real-time operating system (RTOS) designed from scratch.

\subsection{Partitioning Hypervisor}

Our hypervisor relies on hardware-assisted virtualization to achieve efficient
resource partitioning. CPU cores, memory and I/O devices are statically
partitioned during system boot time, which means there is no resource
multiplexing. Each VM is only allowed to access the physical resources within
its domain.


\paragraph{Unified Scheduling.} Although hardware resources are partitioned,
the hypervisor still allows the master OS to indirectly control the resource
usage of vLib OSes. This is achieved by extending the original hypervisor with
our unified scheduling mechanism, which requires coordination between the
master OS, the vLib server and the hypervisor.

Firstly, when the vLib server invokes {\em vLib\_listen}, it jumps into the
hypervisor and blocks the entire VM waiting for vLib calls. Later, a client
thread in the master makes a vLib call into the kernel, which then transfers
control to the hypervisor. A request flag is set to unblock the destination
vLib OS, while input data is passed to it through a pre-created communication
channel. After unblocking, the CPU running the client returns to the master OS
kernel space. Instead of blocking the user thread, the kernel marks it as being
in a {\em remote state} and forces it to busy wait on the channel for a request
completion signal, with interrupt and kernel preemption enabled. This
effectively turns the thread into an idle thread. From the perspective of the
client, the vLib call is a blocking call. However, inside the busy waiting
loop, timestamps are checked to enforce the vLib call timeout. Notice that this
busy waiting approach greatly simplifies the changes that need to be made to
the master OS scheduler, though it leads to a lowered CPU utilization (which
can be avoided, see Section~\ref{sec:disc}).

On the server side, the vLib OS is unblocked and the channel ID is passed to
the vLib server. If the channel has not been mapped before, the server passes
control into the {\em vlibShm} kernel module (details in
Section~\ref{sec:linux}). This module maps the specified communication channel
into the server's address space. After the channel is mapped, the server
commences request handling.

If the waiting client thread runs out of CPU budget, it is descheduled.
Meanwhile, the scheduler generates an Inter-Processor Interrupt (IPI) to the
vLib OS CPU(s). Although normal interrupts are directly delivered to guest OSes
to reduce virtualization overhead, Non-Maskable Interrupts (NMIs) are
configured to cause VM exits. By setting the delivery mode of an IPI to be NMI,
the destination core's control is passed into the hypervisor (similar to how
Jailhouse behaves~\cite{jail}). The hypervisor performs resource accounting
(including cache occupancy and memory bandwidth usage) for the VM which is then
blocked. We call this process {\em remote descheduling}.  All resource usage
data is returned to the master OS and budgeted to the client thread. When the
client thread is dispatched again, an unblock signal is set so that the vLib OS
resumes its execution from where it was previously descheduled. In the
scheduler's view, the client thread is executing the vLib OS service the entire
time. Hence, the execution of client threads and services are unified.

After the server completes a service, it calls {\em vLib\_listen} again. Before
waiting for another request, it sets a completion signal. This allows the
client thread to exit its idle loop, discard its remote state and return to
user space with the service result. Note that in this paper, we focus only on
synchronous vLib calls. We will discuss the issue of asynchronous vLib calls in
Section~\ref{sec:disc}.

\subsection{Real-Time Master OS}

Mostly, we take \SysName{} as it is except a scheduler extension for remote
state handling and a kernel module (vlibShm) for channel mapping. \SysName{}
features a multicore real-time scheduling framework, which combines partitioned
scheduling with dynamic load balancing. Central to the scheduling framework is
the management of virtual CPUs (VCPUs)~\footnote{different from the virtual CPU
concept inside hypervisors}.

\paragraph{VCPU.} VCPUs are created in \SysName{} to serve as resource
containers for corresponding threads. A VCPU accounts for budgeted CPU time
usage for specific threads and serves as an entity against which scheduling
decisions are made. Each VCPU, $V_i$, is specified a CPU budget, $C_i$, and a
period, $T_i$. The system guarantees that a VCPU receives at least its budget
in every period when it is runnable, given the total CPU utilization,
$U=\sum_{i=1}^{n}\frac{C_i}{T_i}$ for $n$ VCPUs, is less than a specific
threshold. In this paper, we assume just a one-to-one mapping between threads
and VCPUs. A thread is bound to a VCPU, which is then scheduled on a core.

A local scheduling queue for each core orders VCPUs using the Rate-Monotonic
Scheduling (RMS) policy~\cite{Liu:73}, which is a static priority preemptive
scheduling algorithm. With this approach, VCPU priorities are inversely
proportional to their periods. RMS has several valuable properties. First, it
provides analyzable bounds on the total CPU utilization of a set of VCPUs,
within which each VCPU's real-time service requirement can be guaranteed.
Second, in overload situations, RMS guarantees service to the highest priority
subset of VCPUs for which there are sufficient resources. RMS analysis shows
that for large numbers of VCPUs, a core guarantees service to a set of VCPUs if
their total utilization does not exceed 69\%. In practice, feasible schedules
are possible for higher utilizations (e.g., if all VCPUs' periods are
harmonically related).

\paragraph{Scheduling Model.} Each VCPU with available budget at the current
time operates in {\em foreground mode}. When a VCPU depletes its budget it
enters {\em background mode}, where it will only be scheduled if there are no
other runnable foreground VCPUs on the same core. A core is said to be in
background mode when all VCPUs assigned to it are in background mode, otherwise
it is in foreground mode. RMS is used only when a core is in foreground mode.
When a core turns into background mode, we take a fair-sharing approach to
scheduling VCPUs. The scheduler attempts to equally distribute the amount of
background mode CPU time (BGT) each VCPU consumes. 

The foreground mode, together with the corresponding CPU reservation, is used
to guarantee an application's base-level service quality, while the use of BGT
is to further improve its progress. This is beneficial to applications that
improve the resolution, or quality, of their results when granted extra
computation time~\cite{anytime:09, Liu:91}. It also provides a performance
lower bound to batch, or other CPU-intensive, workloads. Nevertheless, BGT
should be allocated cautiously in order to avoid excessive cache or memory bus
contention, which will be discussed next.

\begin{figure}[!htb]
\centering
\includegraphics[scale=0.45]{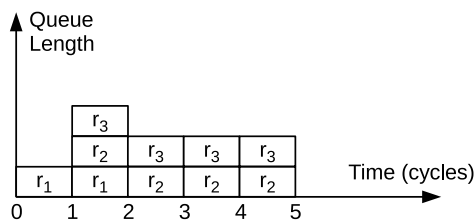}
\caption{Example of Occupancy and Requests}
\label{fig:example}
\end{figure}

\paragraph{Memory Throttling.} Shared memory bus contention on multicore
platforms is one of the major causes of unpredictable application performance,
especially when considering the large volumn of streaming sensor data (e.g.,
3D LIDAR) in modern embedded systems. DRAM accesses on one core might incur
delays due to concurrent accesses on other cores. One approach to address this
issue is to regulate the rate of DRAM references ({\em throttling}), so that
each core cannot exceed a pre-defined bandwidth threshold over a period of
time~\cite{Yun:13}. However, our previous work~\cite{Ye:16} suggests that
rate-based memory throttling has several drawbacks. Instead, we have developed
a latency-based throttling mechanism. Specifically, we measure memory traffic
by directly looking at the average latency to service DRAM memory
requests. Our system uses the PMU to efficiently monitor the average memory
request latency.  Intel Sandy Bridge and more recent processors provide two
uncore performance monitoring events: UNC\_ARB\_TRK\_REQUEST.ALL and
UNC\_ARB\_TRK\_OCCUPANCY.ALL.  The first event counts all memory requests
going to the memory controller request queue ($requests$), and the second one
counts bus cycles weighted by the number of pending requests in the queue
($occupancy$). For example, in Figure~\ref{fig:example}, request $r_1$ arrives
at time 0 and finishes at time 2. $r_2$ and $r_3$ both arrive at time 1 and
complete at time 5. At the end of this 5-cycle period, $occupancy = 10$,
$requests = 3$. We then derive the average latency (cycles) per request as
follows:
\begin{center} 
$latency = \frac{occupancy}{requests}$ 
\end{center}

A bus monitoring thread periodically updates the average latency for the entire
system. Memory throttling starts when the observed average latency hits a
system-configured threshold, which is platform-dependent. Throttling is only
applied to a core in background mode, in which case surplus CPU time is not
allocated to VCPUs and the core is switched to an idle state during its
background mode. This reduces contention on the memory bus and the shared
caches, thus helping VCPUs on other cores make greater progress in their
foreground mode. Instead of simply disabling the cores with the most DRAM
traffic, we adopt a proportional throttling scheme. Basically, cores generating
more traffic are throttled for longer times.

%

\subsection{vLib OS: Linux}
\label{sec:linux}

We use the Ubuntu Server 14.04.5 (4.4.0 kernel) as our single
vLib OS. One CPU core
is dedicated to the Linux vLib OS, while the rest of cores are assigned to
\SysName{}. Virtualization is simplified by applying a patch comprising
approximately 100 lines of code (LOC) to the Linux kernel. This limits Linux's
view of available physical memory, and adjusts I/O device DMA offsets to
account for memory virtualization. To Linux, most of the processor capabilities
are exposed except those associated with VT-x and PMU.

\paragraph{vlibShm.} The design of the vlibShm modules, both in the master OS
and the vLib OS, are very similar to each other, so we focus on our Linux side
implementation for brevity. A communication channel's physical memory is
outside of Linux's memory range and cannot be easily mapped into the vLib
server. The vlibShm kernel module we have developed is specifically designed to
handle this. The server calls {\em mmap} into the module, passing in the
machine physical address of the channel and its memory size, which are returned
by the hypervisor. vlibShm then creates a {\em vm\_area\_struct} object with a
customized page fault handler. When a page ($A$) inside the channel is accessed
for the first time, a page fault occurs and the handler is invoked. This leads
to the allocation of a page ($B$) within Linux's memory range. $B$'s guest
physical address is then passed to the hypervisor through a hypercall, together
with the machine physical address of page $A$. An EPT entry rewrite is
performed in the hypercall so that the channel's page $A$ now maps to a
legitimate memory address ($B$) in Linux.

\subsection{Discussion}
\label{sec:disc}

In this prototype, our goal is to combine the timing predictability of an RTOS
and the rich body of software available on a GPOS. The RTOS hosts the
time-critical and latency-sensitive tasks or control loops, which are required
to ensure the safety of our evaluation platform~\ref{sec:eval}. At the same
time, we take advantage of the abundant commodity software, including vision,
data logging and communication code hosted on Linux. Our system design not only
allows us to combine legacy and custom software, but also enables fine-grained
resource control required to achieve strong performance isolation.

Given our design goals, we adopted a partitioning hypervisor for maximum
predictability and fast I/O manipulation. The downside of this approach is
that, when a vLib OS is not servicing requests its assigned CPUs are unused. We
believe this is an appropriate tradeoff when targeting real-time systems or
general systems with tight tail latency requirements. However, when higher
resource utilization is desired, a traditional hypervisor with hardware
resource multiplexing can be used. Through CPU multiplexing, a vLib call would
cause a VM switch and CPU utilization can be increased. However, we pay the
cost of VM switching, in terms of pipeline stalls, cache and TLB flushing.

In our implementation, a significant development burden has been avoided by
exploiting the vLibOS model. Rather than rewriting the scheduler to manage the
threads associated with vLib OSes side by side with native \SysName{} threads
(or VCPUs), we treat the execution of the former as a special ({\em remote})
state of the latter. As a consequence, the modification to the scheduler in
\SysName{} is minimal ($< 50$ LOC). This shows how easy it is to make an OS
compatible with vLibOS. Admittedly, this approach poses two disadvantages: 1)
running services from $N$ vLib OSes concurrently would require at least $N$
cores dedicated to the master OS; 2) while the client thread is busy waiting,
it cannot yield its CPU before depleting its current budget. The client-side
CPU cannot do additional useful work even if there is no contention on platform
resources. To avoid the utilization issue, one can rewrite the master OS
scheduler to manage different types of threads separately. However, as
demonstrated in our evaluation section, our model affords us the luxury of
doing this only when needed. Our existing platform has sufficient resources to
meet our application goals without this optimization.

We have made a design choice to not support asynchronous vLib calls in
real-time systems. With blocking calls, critical and non-critical code
(including Linux services) have temporal separation within a thread. If an
asynchronous vLib call is made, then the critical code and non-critical code
would be running simultaneously on different cores, impacting each other when
accessing shared resources. The added contention would make it even harder to
guarantee predictable execution on multicore platforms.

While our vLibOS prototype uses a research RTOS as the master and a mature
legacy OS as the slave vLib OS, this need not be the case. In fact, a
full-blown OS such as Linux could be the master as well. Such an architecture
can be exploited to achieve Monolithic kernel decomposition~\cite{Nikolaev:13}.
Similarly it is not necessary for the slaves to be commodity OSes, rather it is
equally viable to construct a runtime in which several specialized slave OSes
are included. For example, specialized new OS kernels~\cite{Belay:14, Peter:14}
can be forked by the master to perform specific and highly optimized tasks
(e.g., network I/O). Mixtures of specialized OSes allow developers to focus
their effort on optimizing one service while delegating other services to
general-purpose OSes. Unlike hybrid systems that lack isolation amongst
kernels~\cite{Park:12, Taku:14}, our approach provides security and performance
isolation for the master OS.

Although our prototyping effort has focused on a particular platform, the
techniques it introduces for controlled interactions between the master and
slaves are applicable to other situations. Achieving higher utilization in data
centers leads to dramatic energy and cost savings~\cite{Verma:15}. Our
architecture provides a VM-based framework to construct cloud runtimes in which
high-priority service applications with stringent QoS requirements are
consolidated with lower-priority batch and best effort workloads on the same
hardware nodes. As is demonstrated in our evaluation, despite using
virtualization, it is possible to achieve precise resource throttling and
isolation between VMs even with respect to low-level resources such as shared
caches and memory buses. For instance, vLibOS can be used to structure a
Xen-based cloud environment where the Dom0 acts as a master OS and is extended
with contention management policies.

%% file: eval.tex
\begin{figure}[!htb]
\centering
\includegraphics[scale=0.25]{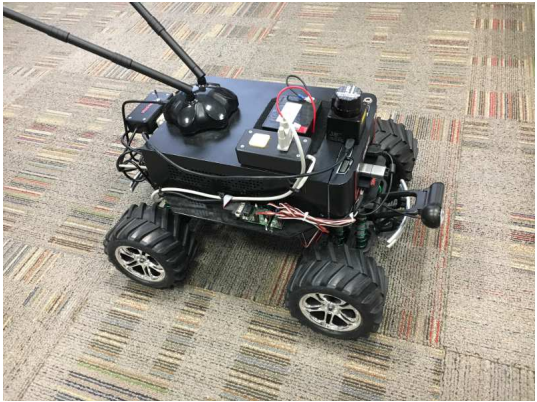}
\caption{Autonomous Vehicle Platform}
\label{fig:car}
\end{figure}

In this section, we evaluate our vLibOS implementation using the hardware
platform as shown in Figure~\ref{fig:car} and Table~\ref{tab:hardware}. The
autonomous ground vehicle houses a custom-made PC. Only three cores are
enabled in the firmware, for the purposes of running all needed tasks and to
conserve energy usage from the main battery powering the vehicle. 
A GeForce GT 710 GPU is used because of its relatively small form-factor,
single PCIE slot requirement, fanless design and low power consumption.

\begin{table}[!htb]
\footnotesize
\begin{center}
\begin{tabular}{| c | c |}
\hline
{\bf Processor} & Intel Core i5-2500k quad-core \\ \hline
{\bf Caches}    & 6MB L3 cache  \\ \hline
{\bf Memory}    & 4GB 1333MHz DDR3 \\ \hline
{\bf GPU}       & MSI GeForce GT 710 2GB \\ \hline
{\bf Camera}    & Logitech QuickCam Pro 9000 \\ \hline
{\bf LIDAR}     & Hokuyo URG-04LX-UG01 \\
\hline
\end{tabular}
\end{center}
\caption{Hardware Specification}
\label{tab:hardware}
\normalsize
\end{table}

We assigned two of the three cores to \SysName{} and one to Linux. Our
hypervisor and \SysName{} relied on an in-RAM file system while Linux used a
USB drive for its storage. Both servo controller and LIDAR (mission-critical
tasks) were connected through serial ports. Based on this system requirement,
we partitioned serial ports to \SysName{} and granted exclusive access of the
GPU and USB stack (camera and storage) to Linux.

\subsection{vLib Call Overhead}

In this experiment, we examined the overhead of the vLib call mechanism. We
started by measuring VM entry/exit costs followed by the cost of making a vLib
call. We ran a test thread in \SysName{} and a vLib server in Linux. The test
thread establishes a communication channel and keeps making vLib calls without
input data. The vLib server receives requests and immediately signals the
completion without performing any services. To avoid the impact of scheduling
in \SysName{}, we measured the time difference $T_{1}$ between when the thread
entered the kernel and when it was about to return to user space. To avoid
Linux scheduling overheads, we measured the time, $T_{2}$, between when {\em
vLib\_listen()} was about to return from the hypervisor (for servicing new
requests) and when the next {\em vLib\_listen()} call entered into the
hypervisor (to generate a completion signal). The vLib call overhead is
represented as $T_{1} - T_{2}$. For the remote descheduling cost, we also
measured the time difference between the moment the \SysName{} kernel sent out
an IPI and when a VM exit was completed on the Linux core. Finally, we also
measured the execution time of our customized page fault handler as the cost
of mapping a single-page communication channel into Linux. All measurements were
averaged over 1000 times and are shown in Table~\ref{tab:cost}. Note that the
channel is mapped only when it is accessed for the first time. It is not on
the critical path of the vLib calls.

\begin{table}[!htb]
\footnotesize
\begin{center}
\begin{tabular}{| c | p{2.3em} | p{1.8em} | p{1.8em} | p{3.5em} | p{3.7em} |}
\hline
            & VM Entry    & VM Exit     & vLib Call    & Remote Desched   & Channel Mapping \\ \hline
CPU Cycles  & 531         & 481         & 4754          & 1153            & 2377 \\
\hline
\end{tabular}
\end{center}
\caption{Mechanism Overhead}
\label{tab:cost}
\end{table}
\normalsize

\subsection{Performance of Partitioned I/O Devices}

Our partitioning hypervisor incurs minimum overhead for guest I/O operations.
To measure this overhead, we evaluated the GPU performance in the Linux vLib OS
using an open source neural network application, Darknet~\cite{darknet}.
Although there are other popular deep learning frameworks available, we chose
Darknet because of its support for 32-bit platforms. Darknet is implemented in
C and CUDA with high efficiency and only a few dependencies. We believe it is
well suited to embedded applications.

In this experiment, we compared the performance of running Darknet in the
stand-alone Linux (vanilla Linux) and Linux running on top of \VirtName{} (vLib
Linux). Both Linux kernels were built without SMP support for fair comparison.
The vLib Linux still retains execution control over its CPU since the vLib
server was not started for this experiment. We measured the execution time of
the Darknet image classification operation (CUDA code) on the GPU for both
systems. The results were averaged over 1000 operations on the same single
image and are shown in Table~\ref{tab:io}. As can be seen, the vLib Linux
achieved similar GPU performance ($7\%$ slowdown) comparing to the vanilla
Linux. Notice that part of the slowdown comes from the memory virtualization
overhead.

\begin{table}[!htb]
\footnotesize
\begin{center}
\begin{tabular}{| c | c | c | c |}
\hline
vanilla Linux   & 859       & vLib Linux      & 920 \\
\hline
\end{tabular}
\end{center}
\caption{GPU Performance ($10^6$ CPU cycles)}
\label{tab:io}
\normalsize
\end{table}

\subsection{Effectiveness of Memory Throttling}
\label{sec:jump}

To evaluate the effectiveness of the \SysName{} memory throttling mechanism,
we introduced a memory-intensive micro-benchmark, {\em m\_jump}, to measure
the memory bus performance (Listing~\ref{code:mjump}). It operates on a 6MB
data array, which is large enough to occupy the entire L3 cache. The benchmark
writes to the first 4 bytes of every 64 bytes in the array. As every cache
line is 64 bytes, this causes the entire cache to be filled. After every
write, {\em m\_jump} jumps 8KB forward in order to avoid reusing data from
DRAM row buffers~\cite{Ye:16}. It is worth noting that caches cannot be
disabled for this experiment, even though our focus is on memory bus
performance. If caches were disabled, every instruction needs to be fetched
from memory, effectively forcing CPUs to run at the same speed as the memory
bus and reducing the likelihood of bus congestion.

\footnotesize
\begin{lstlisting}[caption={\em m\_jump}, label=code:mjump]
      byte array[6M]; 
      for (uint32 j = 0; j < 8192; j += 64)
         for (uint32 i = j; i < 6M; i += 8192)
            array[i] = i;
\end{lstlisting}
\normalsize


We set up three groups of experiments, each with a different CPU foreground
utilization (C/T) for {\em m\_jumps}. In each group, we ran five 10-minute
experiments for comparison. The first experiment ({\em alone}) ran a single
{\em m\_jump} under \SysName{}, without a co-runner. At the end of the
experiment, we recorded {\em m\_jump}'s instructions retired only in the
foreground mode ({\em FG Inst}). For the second ({\em \SysNameS{}}) and the
third ({\em \SysNameS{} + mem}) experiments, two {\em m\_jumps} were started at
the same time on different cores in \SysName{}. We disabled memory throttling
for the second experiment and enabled it for the third. We measured the {\em FG
Inst} for the first of the two {\em m\_jumps} for both experiments. In the
fourth experiment ({\em linux}), the first {\em m\_jump} was started in \SysName{}
while the second was started in Linux. Memory throttling was enabled in
\SysName{}. The performance of the {\em m\_jump} in \SysName{} was measured.
The last experiment ({\em linux + mem}) was similar to the fourth, but the {\em
m\_jump} in Linux was invoked by a \SysName{} thread through a single vLib call
(timeout set to null) so that memory throttling could be applied to Linux.

\begin{figure}[!htb]
\centering
\includegraphics[scale=0.7]{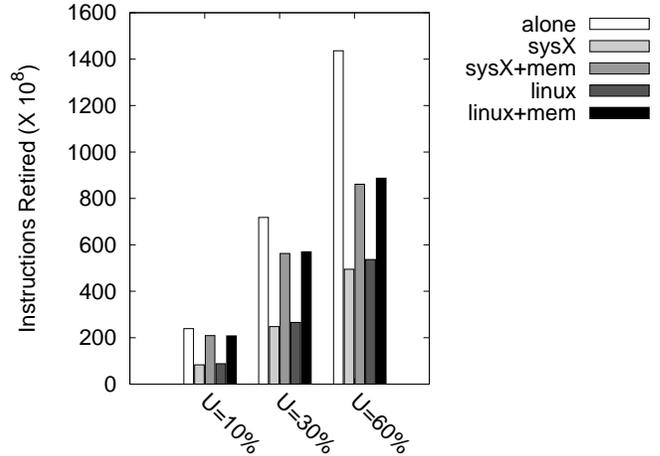}
\caption{Effective Memory Throttling}
\label{fig:jump}
\end{figure}

In group one (U=10\%), the first {\em m\_jump} was bound to a VCPU with C=10,
T=100, where the time was in milliseconds. For the second {\em m\_jump}, C=9,
T=90. We refer to this setting as \{C=(10, 9), T=(100, 90)\}. The CPU
utilization of both threads were 10\% but the periods ($T$s) were set
differently to reduce the likelihood of memory accesses occurring all
together~\cite{Ye:16}. When running {\em m\_jump} in Linux, there was no VCPU
assignment. The VCPU with C=9 and T=90 was assigned to the \SysName{} thread
performing the vLib call in the fifth experiment. For group two (U=30\%) and
group three (U=60\%), VCPU settings were \{C=(30,27), T=(100, 90)\} and
\{C=(60,54), T=(100, 90)\}, respectively. Figure~\ref{fig:jump} shows the {\em
FG Inst} of the first {\em m\_jump} in different cases.

Comparing the {\em linux} case to the {\em alone} base case, we see that having
uncontrolled bus contention inside Linux leads to a significant performance
drop for real-time threads running in \SysName{}. If memory throttling is
applied to Linux using our unified scheduling mechanism, as in {\em linux +
mem}, a large reduction in performance slowdown is achieved. In group U=10\%,
there is a $50\%$ reduction in slowdown.

As we increase the CPU utilization of {\em m\_jump}, its foreground performance
increases correspondingly, due to increased foreground time. However,
background time decreases at the same time. With less background time, the
effectiveness of memory throttling is reduced because there is less slack time
to stall execution on individual cores.

\subsection{Autonomous Driving Case Study}

\begin{figure}[!htb]
\centering
\includegraphics[scale=0.4]{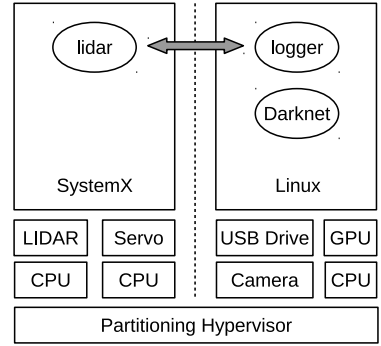}
\caption{Autonomous Driving System}
\label{fig:system}
\end{figure}

We then tested \VirtName{} using a real-time application involving an
autonomous ground vehicle. This system (see Figure~\ref{fig:system}) consists
of a real-time program {\em lidar} and two Linux programs, logger and Darknet.
{\em lidar} takes LIDAR data as input and makes steering decisions to avoid
objects. Meanwhile it dumps data to a 4MB memory buffer, which is shared with
the logger. The logger periodically checks the buffer and saves data to a log
file in the USB drive when the buffer is full. Saved data is used for offline
diagnostics. Darknet runs side by side in Linux, reading camera frames and
performing object classification, which is useful in autonomous vehicle
control. For example, Darknet is able to identify traffic signs while LIDAR is
not. 

We considered the {\em lidar} as the most critical part of the system, so it
was placed in \SysName{}. Data logging and object classification improve
quality of service, but their failure is tolerable as long as obstacle
avoidance is working. Although less critical, implementing them from scratch
would take significant engineering effort. Using a legacy Linux implementation
of Darknet, with data logging, greatly reduced the time to build our
mixed-criticality system.

\begin{figure}[!htb]
\centering
\includegraphics[scale=0.4]{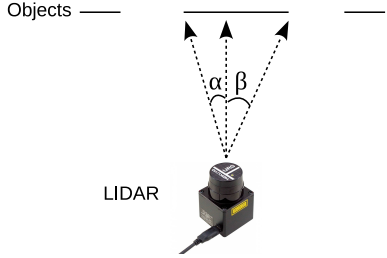}
\caption{Object Avoidance Algorithm}
\label{fig:avoid}
\end{figure}

In our object avoidance solution, the LIDAR device sends out distance data
(around 700 bytes) every 100 ms. Periodically, {\em lidar} decodes the received
data and scans object distances from all angles (240 degrees) in order to
identify objects within a certain distance. If it finds a nearby object
directly in front of the vehicle, it then looks for the closest open space
either on the left side or on the right side. From the example in
Figure~\ref{fig:avoid}, since $\alpha < \beta$, a left turn decision will be
made.

In our evaluation, we divided experiments into 3 cases. In the first one ({\em
lidar}), {\em lidar} was running in \SysName{} with a VCPU configured with
\{C=12, T=40\}. Although the logger was started in Linux as well, during the
whole experiment time, the shared buffer did not fill to capacity. Thus, the
logger did not perform any task. We consider this case as {\em lidar} running
alone on the platform. This allows us to focus on evaluating only the
performance impact from co-running Darknet later. We will not mention the
logger again in the experiment description that follows. 

In every period, we measured the execution time of {\em lidar} from right
after it received LIDAR data to the end of its object avoidance
algorithm. 2000 samples were taken during the experiment. In the second case
({\em lidar+Darknet w/o mem}), we simultaneously ran {\em lidar} in \SysName{}
and Darknet in Linux. In Linux, we did not start a vLib server, so Linux still
had control over its own execution. The same measurement was carried out. The
last experiment ({\em lidar+Darknet w/ mem}) was similar to the second, except
a vLib server ran in Linux and \SysName{} acquired full system
control. A \SysName{} thread ({\em client}) was created with VCPU \{C=12,
T=40\}. Immediately after starting, it made a blocking vLib call to Linux with
timeout set to null. Without sending out a request completion signal, the
server simply terminated itself. Darknet then kept running inside Linux with
the {\em client}'s CPU budget, so that memory throttling could be applied.


\begin{figure}[!htb]
\centering
\includegraphics[scale=0.4]{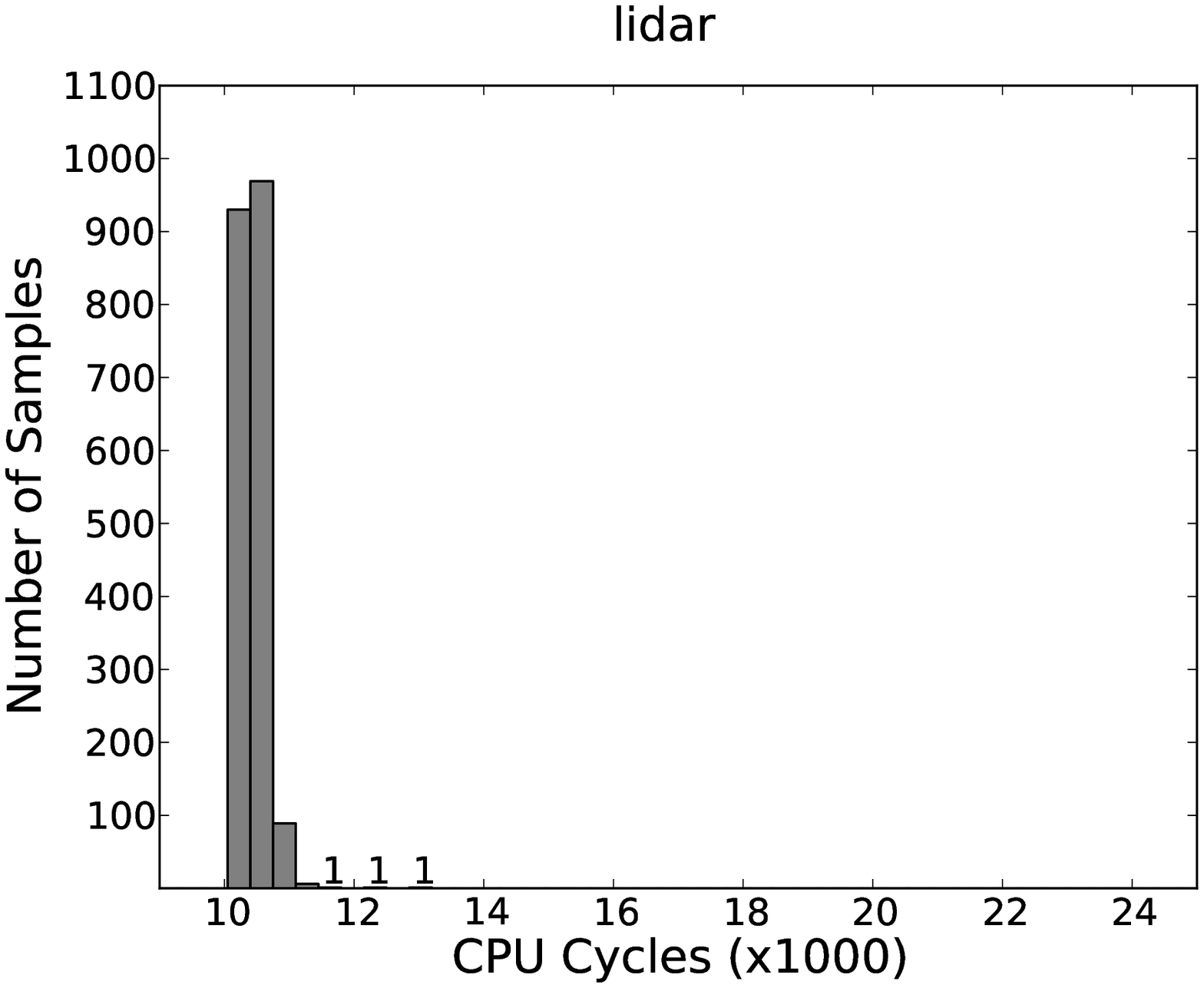}
\includegraphics[scale=0.4]{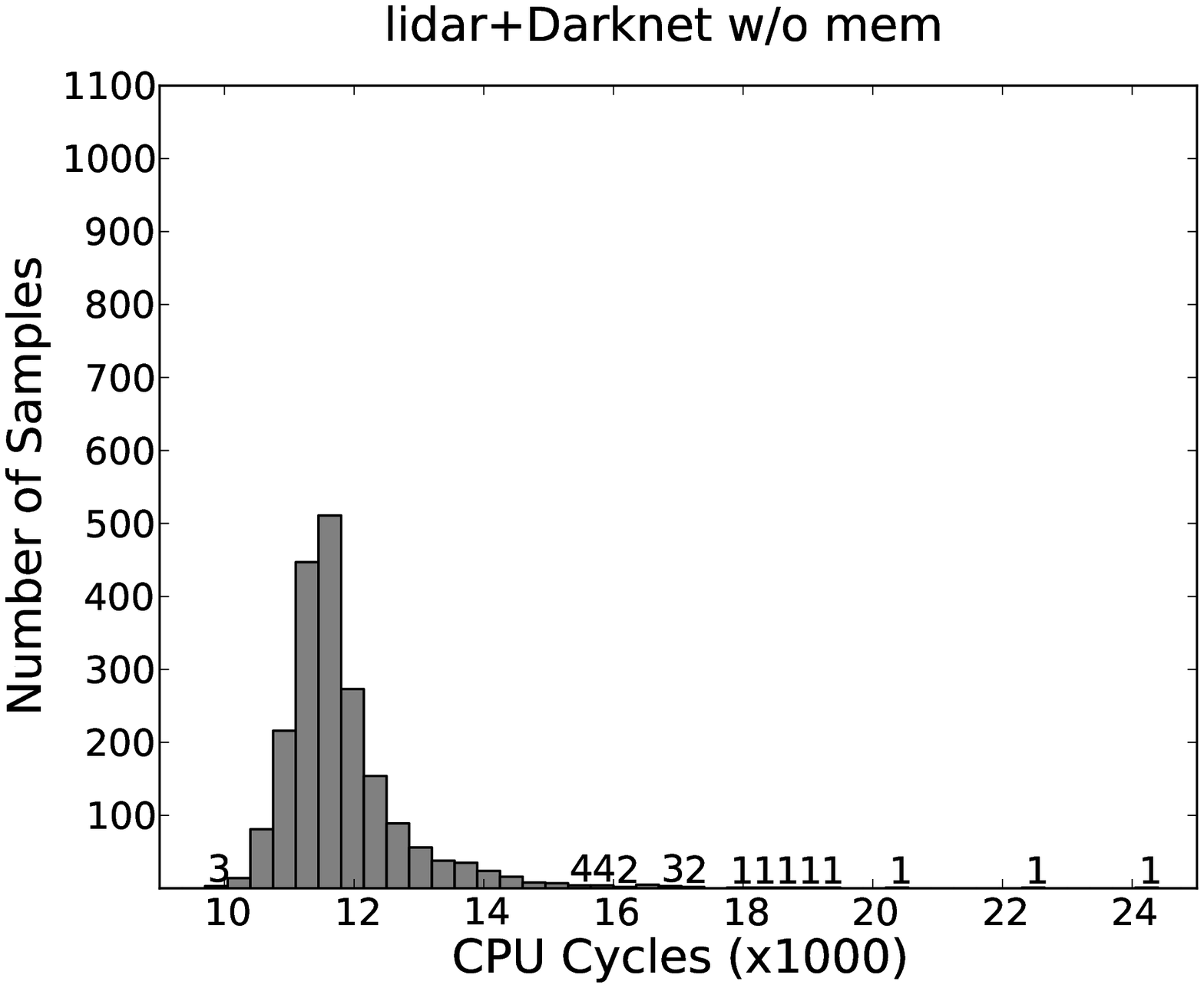}
\includegraphics[scale=0.4]{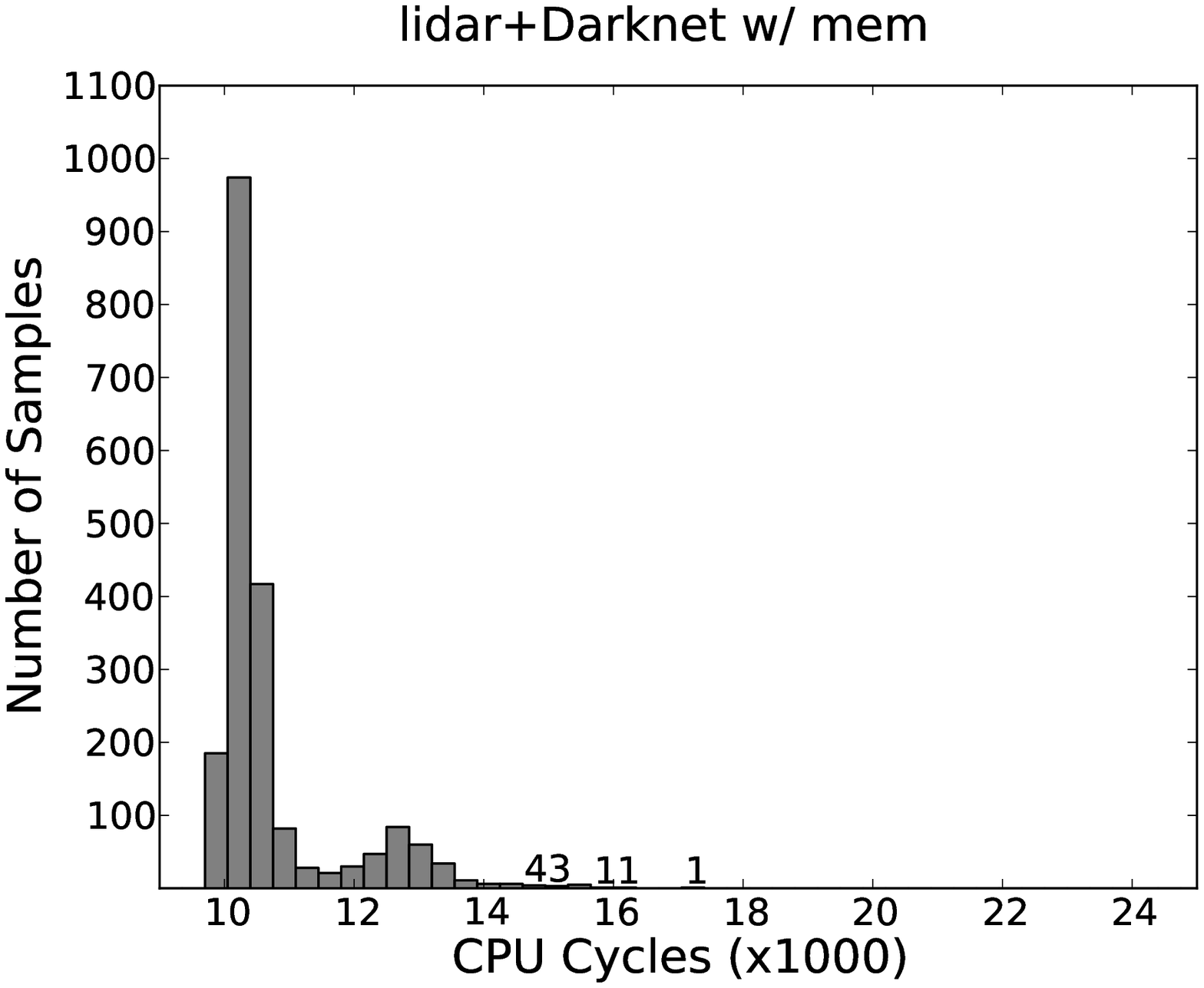}
\caption{{\em lidar} Performance in \VirtName{}}
\label{fig:lidar}
\vspace{-0.15in}
\end{figure}

From Figure~\ref{fig:lidar}, we can see that {\em lidar}, when running
alone inside \SysName{}, has a very stable performance. When Darknet starts
competing for shared resources in case {\em lidar+Darknet w/o mem}, {\em
lidar} suffers from increased performance variation. The worst case execution
time we observed was around 24000 CPU cycles, which is twice the average. This
is because Linux was not running a vLib server and could not be controlled
by \SysName{}. When the vLib call mechanism was enabled in case {\em
lidar+Darknet w/ mem}, memory bus contention was effectively managed, leading
to reduced performance variation. The worst case execution time dropped to
around 17000 cycles.

It is worth mentioning that the lidar program's working set fits into the
private L1/L2 caches. This explains why the {\em lidar}'s average execution
time does not increase as much as we would expect from the previous
section~\ref{sec:jump}, in the presence of memory bus contention. However, more
advanced LIDAR devices, better object avoidance algorithms and more complicated
sensor fusion algorithms all contribute to a larger memory footprint of a
real-time program. As a result, memory bus management would become essential.

For comparison, we also investigated how a vanilla Linux would perform on our
autonomous vehicle. Ideally, we would use the RT-PREEMPT patch for Linux,
which improves performance for real-time tasks. However, the Nvidia GPU driver
did not support the real-time patch, so we were restricted to using an
unpatched SMP Linux system.

\begin{figure}[!htb]
\vspace{-0.15in}
\centering
\includegraphics[scale=0.4]{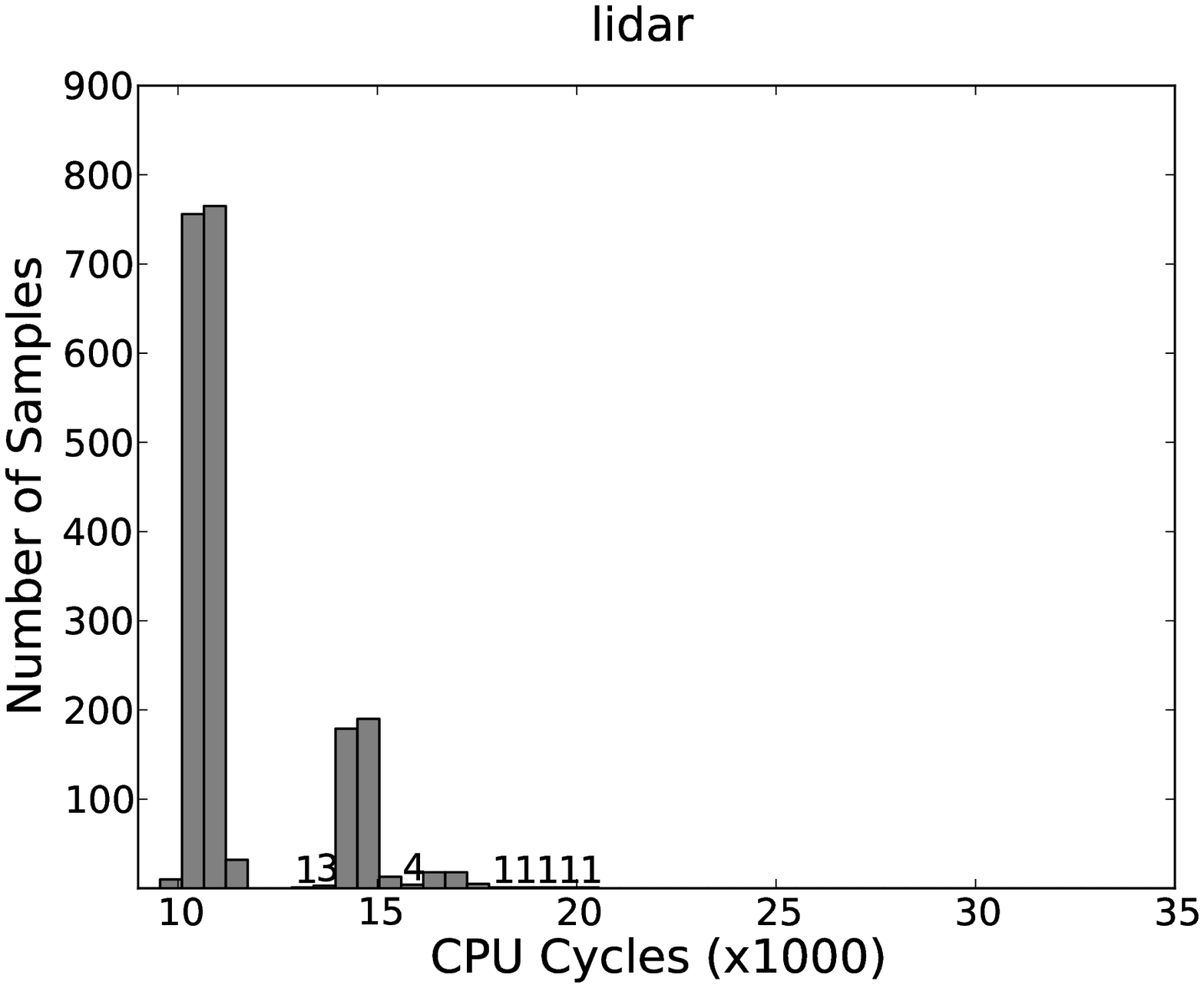}
\includegraphics[scale=0.4]{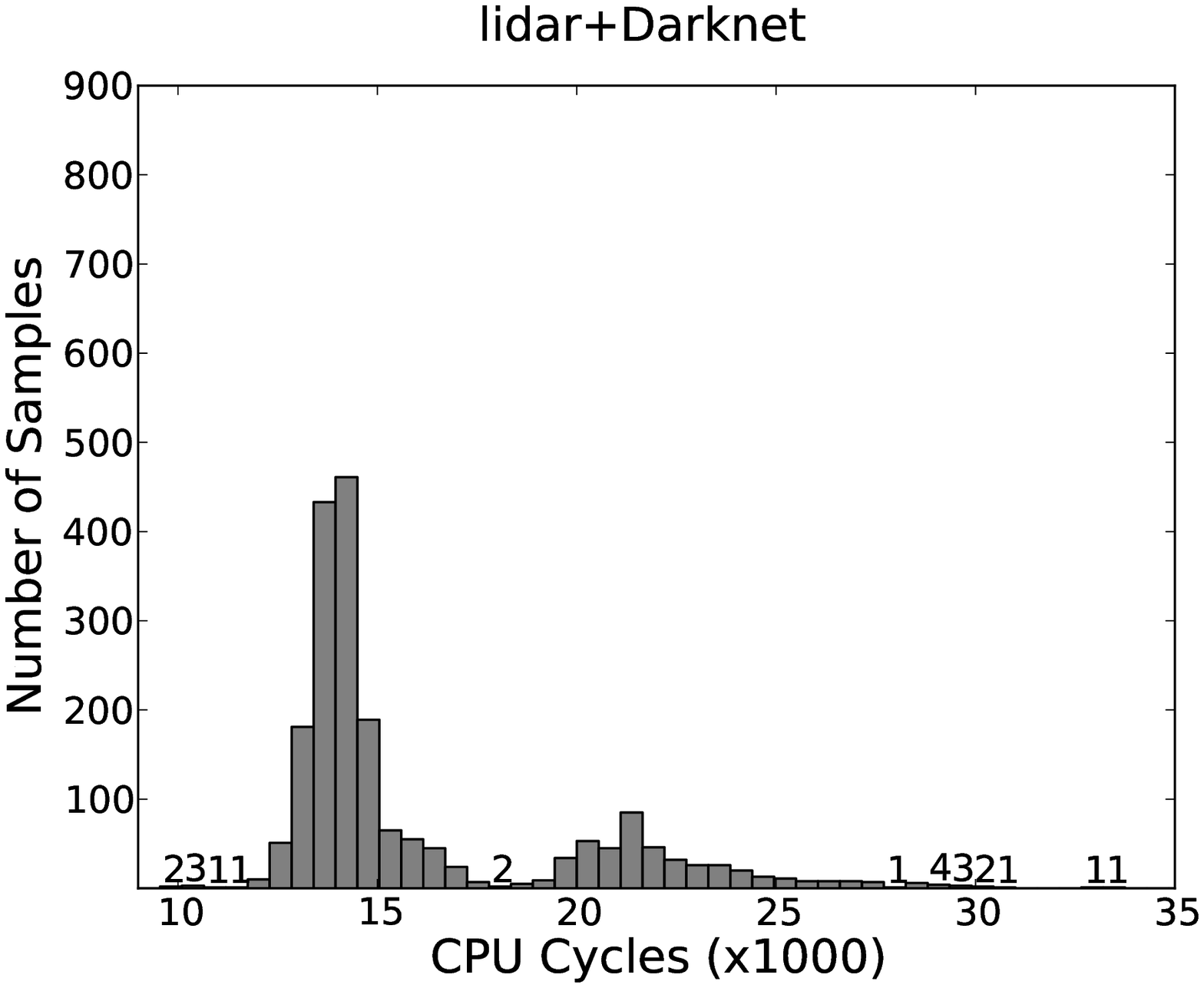}
\caption{{\em lidar} Performance in Vanilla Linux}
\label{fig:linux}
\vspace{-0.1in}
\end{figure}

For the first experiment ({\em lidar}), we pinned {\em lidar} to a core
together with a CPU hog, which runs an empty while loop. Since {\em lidar}
itself runs only periodically, it does not create much workload for the
CPU. If we do not assign a hog on the same CPU, Linux performs Dynamic Voltage
and Frequency Scaling (DVFS) on the CPU, thereby decreasing its
frequency. This unnecessarily slows down the execution of {\em lidar} and
impacts our measurements. We also set the real-time scheduling class
SCHED\_FIFO to {\em lidar} with the highest priority in order to avoid
preemption. The execution time of {\em lidar}'s periodic task was measured
over 2000 samples. Next, we put Darknet on another CPU and repeated the same
experiment ({\em lidar+Darknet}). Results are shown in Figure~\ref{fig:linux}.

As Linux is not designed for real-time applications, {\em lidar} experienced
noticeable performance variation even running alone. With the presence of
interference by Darknet, both the average and worst case execution time were
prolonged significantly. Comparing these results with the results from
Figure~\ref{fig:lidar}, we believe \VirtName{} provides better real-time
service to the lidar application.

%% file: related.tex
\subsection{OS Evolution Strategies}

OSKit~\cite{Ford:97} is an early work with an explicit goal to ease new OS
development. It provides a set of commonly used OS components, like
bootstrapping, architecture-specific manipulation and protocol stacks.
Modularization and encapsulation of legacy code via a glue layer allow
developers to concentrate their engineering effort on innovative features.
However, it is hard to achieve comparable performance with existing systems if
strict modularization is used. Also, creating an adaptation layer may not be an
easy task.

There have been a number of research efforts focusing on OS structure and
extensibility. Extensible operating systems research~\cite{small96comparison}
aims at providing applications with greater control over the management of
their resources.  For example, the exokernel~\cite{Engler:95} tries to
efficiently multiplex hardware resources among applications that utilize
library OSes.  Resource management is thus delegated to library OSes, which
can be readily modified to suit the needs of individual applications.
SPIN~\cite{Bershad:SPIN} is an extensible operating system that supports
extensions written in the Modula-3 programming language.  Interaction between
the core kernel and SPIN extensions is mediated by an event system, which
dispatches events to handler functions in the kernel. By providing handlers
for events, extensions can implement application-specific resource management
policies with low overhead.

Exokernels, library OSes like Drawbridge~\cite{Porter:11} and
unikernels~\cite{Madhavapeddy:13} all view OS services as a set of application
libraries. vLibOS differs from them by treating an entire legacy OS as a
single library whose execution is mediated by a master OS. With vLibOS there
is no need to reimplement a legacy OS as a set of library services.

Some work have attempted to virtualize existing OSes, which offer services to a
new OS. User-Mode Linux~\cite{UML} and L4Linux~\cite{l4linux}, for example,
implement a modified Linux inside a user-level address space on top of a host
OS. Libra~\cite{Ammons:07}, on the other hand, relies on virtualization
technologies for hosting unmodified guest services. EbbRT~\cite{Schatzberg:16}
employs a similar approach, enabling kernel innovation in a distributed
environment. In EbbRT, services are offloaded between machines running
light-weight kernels and full-featured OSes. Offloading is facilitated by an
object model that encapsulates the distributed implementation of system
components.

Dune~\cite{Belay:12} uses hardware virtualization to expose privileged
hardware features to user-level processes, improving efficiency for certain
applications such as garbage collection. As with extensible kernels, the aim
is to enrich the functionality within existing systems. In contrast, vLibOS
helps develop custom OSes that implement new services based on legacy
functionality in separate VMs.

VirtuOS~\cite{Nikolaev:13} delegates part of its services to other VMs through
an exception-less system call mechanism~\cite{Soares:10}. The execution of
services is controlled by each service domain instead of the primary
domain. An exception-less system call can be implemented on top of our vLib
call. However, VirtuOS, as with Nooks~\cite{Swift:03}, focuses on safe
isolation of existing system components rather than recycling legacy
components for new OSes, as covered in this paper. That being said, vLibOS is
able to restructure the functionality of monolithic kernels across separate
protection domains.

FusedOS~\cite{Park:12} proposes the use of a full-blown OS as a master OS,
which spawns light-weight kernels to a subset of CPUs, but without
virtualization. Shimosawa et al.~\cite{Taku:14} further formalize this hybrid
kernel design and define a corresponding interface. Developers are able
to implement new features in light-weight kernels, while requesting legacy
services through cross-kernel service delegation. The downside is that, without
virtualization, protection between kernels is not enforced. Also, a
light-weight kernel does not have the ability to manage global resources in
order to maintain its desired QoS, as is done in vLibOS.

Commercial embedded systems like PikeOS~\cite{pikeos} and QNX~\cite{qnx} rely
on their own in-kernel virtualization technologies to support legacy
software. However, contention for shared hardware resources is not addressed.
While the RTS Hypervisor~\cite{rts} throttles the memory throughput for
non-critical VMs, it does so using a demonstrably inferior rate-based
(bandwidth) threshold, rather than a latency-based threshold used in our
system. Finally, Litmus$^{RT}$~\cite{litmus} provides extensions to Linux to
support prototype development of real-time schedulers and synchronization
protocols. However, it does not focus on the enforcement of temporal and
spatial isolation between custom and legacy components, as addressed by
vLibOS.

\subsection{Multicore Resource Management}

The effects of shared caches, buses and DRAM banks on program execution have
been studied in recent years. Page coloring~\cite{Bugnion:96,Sherwood:99} is a
commonly used software technique to partition shared caches on multicore
processors. Tam et al.~\cite{Tam:07} implemented static cache partitioning with
page coloring in a prototype Linux system, improving performance by reducing
cache contention amongst cores. COLORIS~\cite{Ye:14} demonstrated an efficient
method for dynamic cache partitioning, enhancing system QoS. In the meantime,
hardware vendors developed their own cache protection mechanisms (e.g., Intel
CAT, ARM cache lockdown) so that caches are better managed.

In terms of memory bus contention, Blagodurov et al.~\cite{Blagodurov:10}
identified it as one of the dominant causes of performance slowdown on
multicore processors and tried to avoid running memory intensive applications
concurrently. Later, MemGuard~\cite{Yun:13} was developed to control memory
bandwidth usage across different cores. Each core is assigned a memory budget,
which limits the number of DRAM accesses in a specified interval. To improve
bandwidth utilization, MemGuard predicts the actual bandwidth usage of each
core in the upcoming period. For cores that do not use all their budgets, they
contribute their surplus to a global pool, which is shared amongst all cores.
PALLOC~\cite{Yun:14} is another approach that uses a bank-aware memory
allocator to assign page frames to applications so that DRAM bank-level
contention is avoided.

Dirigent~\cite{Dirigent:16} is a system that regulates the progress of
latency-sensitive programs in the presence of non-latency-sensitive programs.
It reduces performance variation of specific applications in the presence of
memory contention.  The system works by first offline profiling the execution
of latency-sensitive programs when running alone. An online execution time
predictor/controller then adjusts resources available to them during the
normal runs, to ensure their latency constraints in the presence of
contention. This contrasts with vLibOS's way of throttling cores to avoid
memory contention.